\documentclass[aps,prl,twocolumn,floatfix,superscriptaddress]{revtex4}

\usepackage{epsfig}
\usepackage{amsmath}

\begin{document}

\title{Elliptic flow of thermal photons and formation time of
 quark gluon plasma at RHIC}
\author{Rupa Chatterjee}   
\affiliation{Variable Energy Cyclotron 
Centre, 1/AF Bidhan Nagar, Kolkata 700 064, India}            
\author{Dinesh K. Srivastava}
\affiliation{Variable Energy Cyclotron 
Centre, 1/AF Bidhan Nagar, Kolkata 700 064, India}            
\date{\today}

\begin{abstract}
We calculate the elliptic flow of thermal photons from Au+Au collisions at
RHIC energies for a range of values for the formation time $\tau_0$ but
a fixed entropy (or particle rapidity density). The results are found to 
be quite sensitive to $\tau_0$. The $v_2$ for photons decreases as 
$\tau_0$ decreases and admits a larger contribution from the QGP 
phase which has a smaller $v_2$. The elliptic flow coefficient for 
hadrons, on the other hand, is only marginally dependent on $\tau_0$.
\end{abstract}

\pacs{25.75.-q,12.38.Mh}
\maketitle
It is by now generally believed that we are witnessing the birth
of quark gluon plasma in relativistic collision of gold nuclei at RHIC
energies. This view is supported by the clear display of jet-quenching
~\cite{jetq_theo,jetq_exp}, elliptic flow of particles in non-central 
collisions~\cite{v2_theo,v2_exp} heralding the on-set of collectivity 
and conversion of initial spatial anisotropy to azimuthal anisotropy 
of their momenta, and radiation of thermal photons~\cite{phenix_phot}.  
Now the next set of questions occupy the central stage in this endeavour. 
What is the temperature of the plasma? Is it in thermal and chemical 
equilibrium? What is the formation time $\tau_0$, beyond which we could 
possibly use the powerful methods of hydrodynamics to describe the 
evolution of the system?   

The question of $\tau_0$ is also related to the process of 
thermalization. In the simplest treatment one assumes that the 
partons produced in the collision have an average energy $\langle 
E \rangle$ and thus  $\tau_0\approx 1/\langle E \rangle$. It is a 
different question whether these are thermalized~\cite{kms}. One 
may use one of the many mini-jet based models and assume that the 
energy and momentum of the partons are "some-how" redistributed and 
the plasma is formed at the deposited energy density~\cite{biro}. A 
further refinement may include use of a parton cascade model
~\cite{kkg,bms,xu}, a self-screened parton cascade model~\cite{sspc}, 
or parton saturation models~\cite{eskola} to get a time $\tau_0\approx 
1/p_T$ or $\alpha_s/Q_s$ (Ref.~\cite{son}) at which one assumes the 
plasma to be thermally equilibrated. The chemical equilibration can 
also be modeled using master equations~\cite{biro,munshi}. Even 
though discussed in literature for quite some time~\cite{stan}, the 
role of plasma instabilities in thermalizing the plasma is being 
explored in detail only now (see Ref.~\cite{guy_inst}).  

In view of the several treatments and  perhaps complimentary models 
available in the literature, a more direct measurement of the formation  
time $\tau_0$ would be very desirable. In the present work we argue that 
a experimental determination of the $v_2$ of thermal photons for non-
central collisions can help us determine this value rather accurately.  

\begin{figure}[ht]
\centerline{\epsfig{file=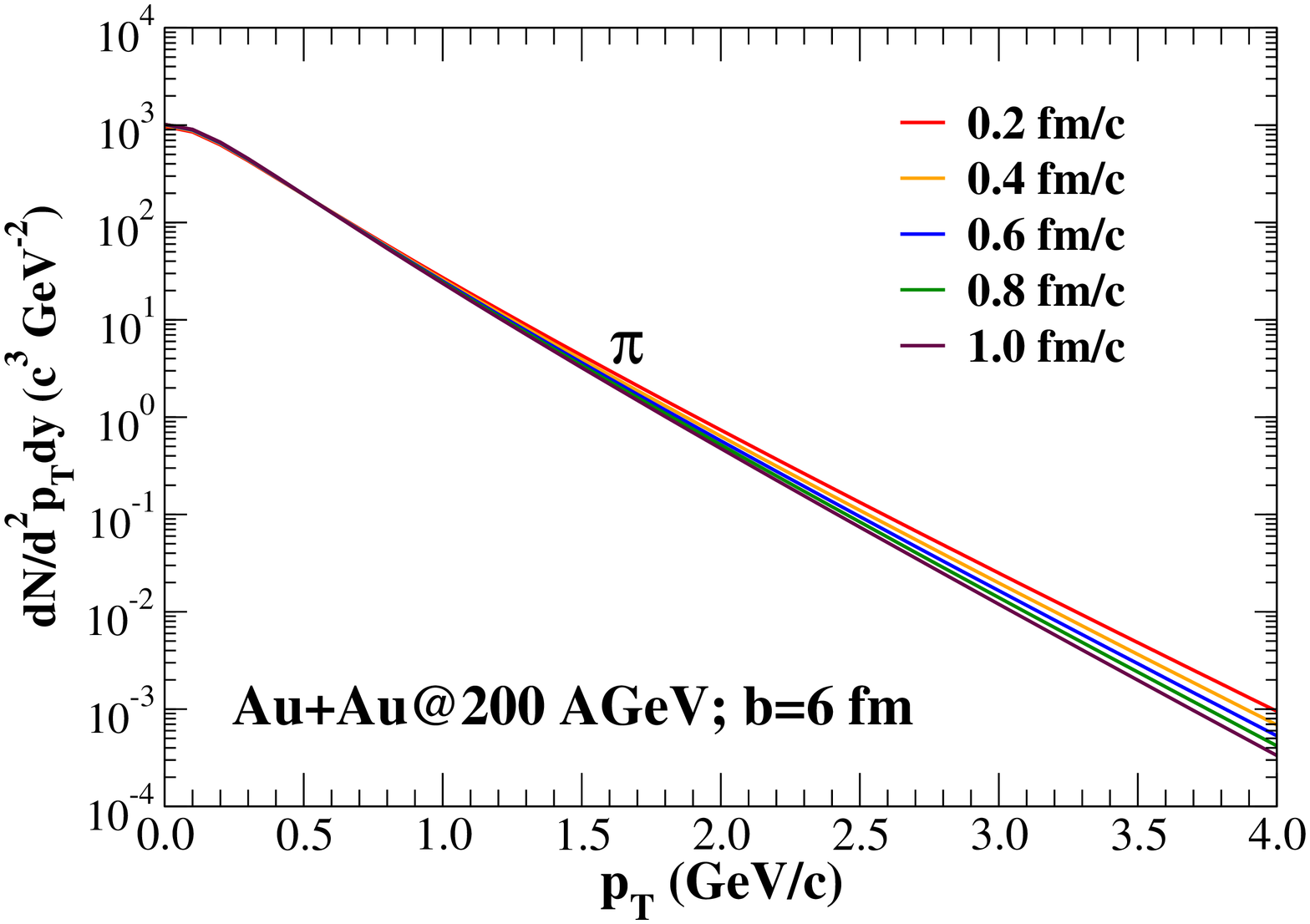,angle=-90,width=7.9cm}}
\centerline{\epsfig{file=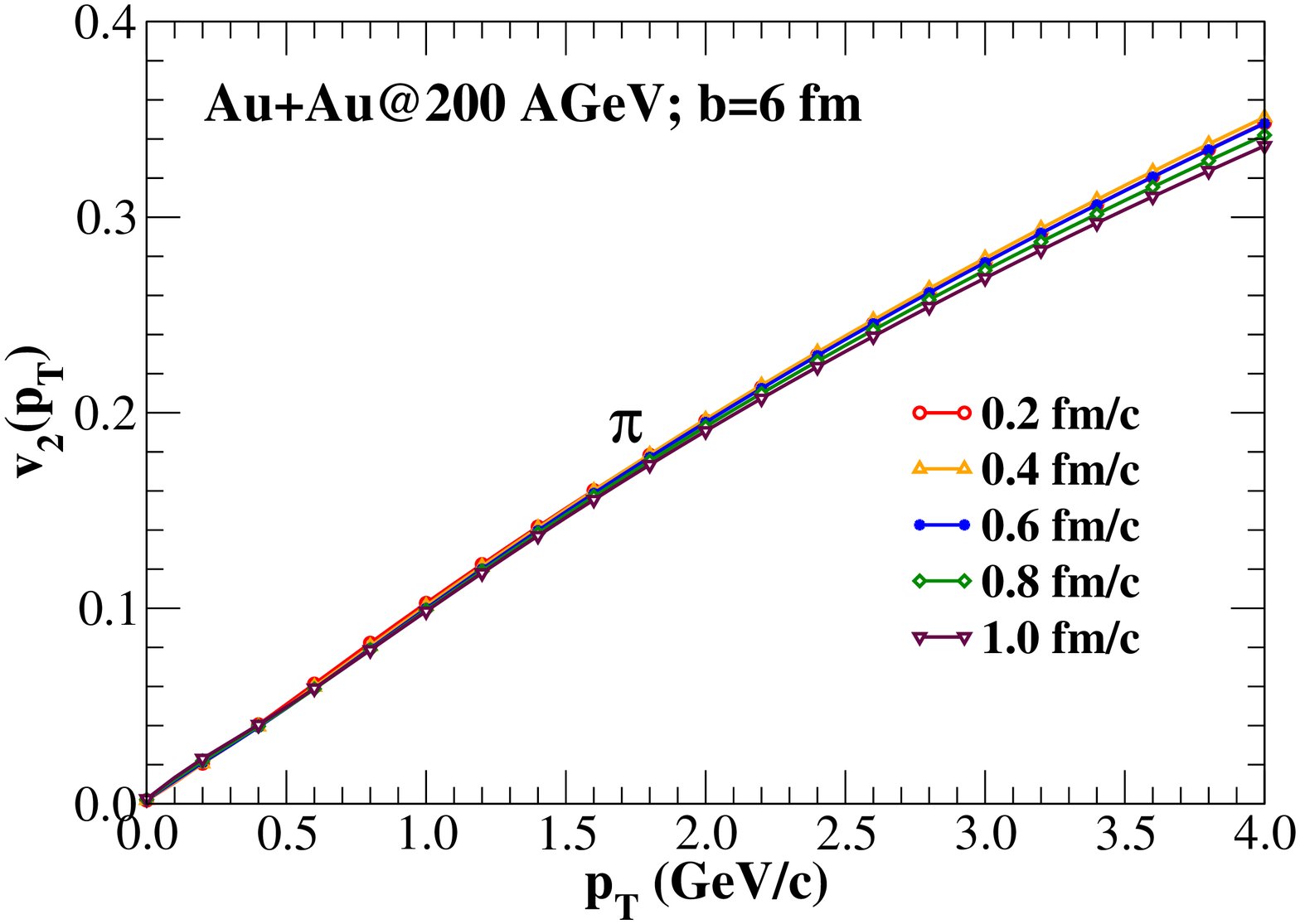,angle=-90,width=7.9cm}}
\caption{(Colour on-line) Spectrum (upper panel)  and $v_2$ (lower panel) 
for primary pions from Au+Au collisions at 200 GeV/A for $b=$ 6 fm for 
different initial times.}
\label{fig1}
\end{figure}

Elliptic flow of thermal photons and dileptons~\cite{v2_phot,v2_dil} has 
recently been proposed as a powerful tool for determining the azimuthal
anisotropy of momenta of partons soon after the formation of the quark-
gluon plasma. The $v_2$ for hadrons though seeded in the early QGP phase 
(as suggested by the re-combination models~\cite{bass} and quark-number 
scaling of hadronic $v_2$) determines this indirectly.

We start with the assumption that a chemically and thermally equilibrated
quark gluon plasma is produced at time $\tau_0$ beyond which we trace its 
evolution using an ideal hydrodynamics. We closely follow the procedure 
outlined earlier~\cite{v2_phot,v2_dil} in setting up the initial conditions 
and assume that the entropy density $s(\tau_0,x,y,b)$ is given by:
\begin{equation}
s(\tau_0,x,y,b)=\kappa\,\left[\alpha \, n_w(x,y,b)+(1-\alpha)\, n_b(x,y,b)
\right]
\end{equation}
where $\kappa$ is a constant, $n_w$ is number of wounded-nucleons 
(participants) at impact parameter $b$, and $n_b$ is the number of 
binary collisions estimated using the Glauber model. In addition, $\alpha$ = 
0.75 is a constant which controls the relative contributions of soft and 
hard processes. We choose $\kappa$ such that $s_0=s(\tau_0=0.6~{\rm~{fm}},
~x=0,~y=0,~b=0)$= 117 fm$^{-1}$ for the collision of gold nuclei~\cite{uk} 
at $\sqrt{s}=$ 200 A GeV. We adjust $\kappa$ to get the starting values 
for $s_0$ at other values of $\tau_0$ such that $s_0 \tau_0$ (or
the particle rapidity density~\cite{bj}) is a constant, which implies 
identical entropy for all the cases. Note that the profile of the 
entropy distribution is uniquely determined by the above equation. 
We employ an impact parameter dependent, boost invariant  hydrodynamics 
\cite{ksh}, which has been used extensively to explore hadron production 
and elliptic flow of hadrons as well as photons~\cite{v2_phot} and 
dileptons~\cite{v2_dil}. 

\begin{figure}[ht]
\centerline{\epsfig{file=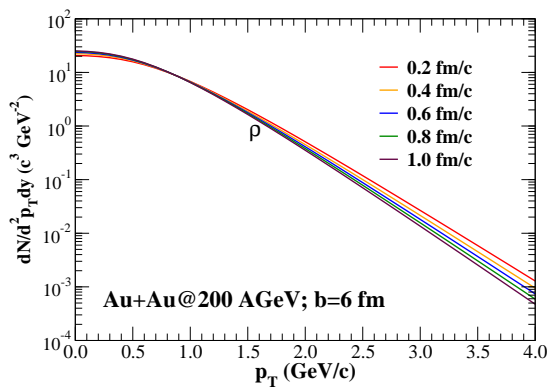,angle=-90,width=7.9cm}}
\centerline{\epsfig{file=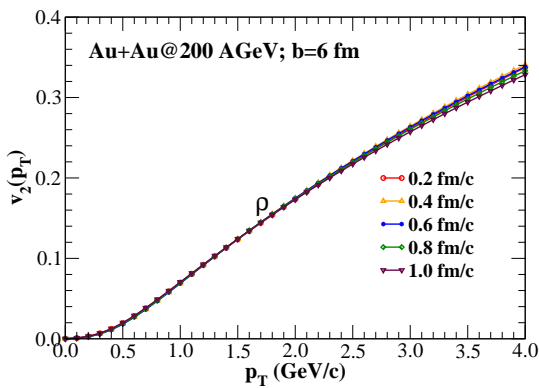,angle=-90,width=7.9cm}}
\caption{(Colour on-line) Same as Fig.~\ref{fig1} for rho-mesons.}
\label{fig2}
\end{figure} 

\begin{figure}[ht]
\centerline{\epsfig{file=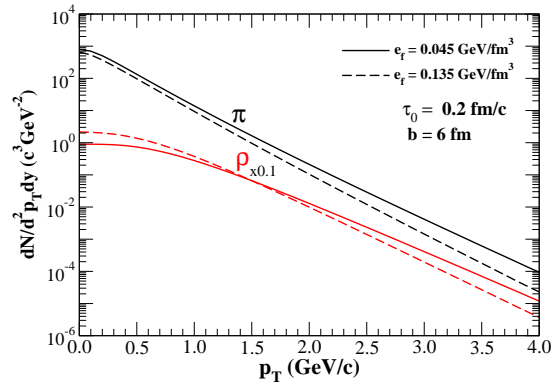,angle=-90,width=7.9cm}}
\centerline{\epsfig{file=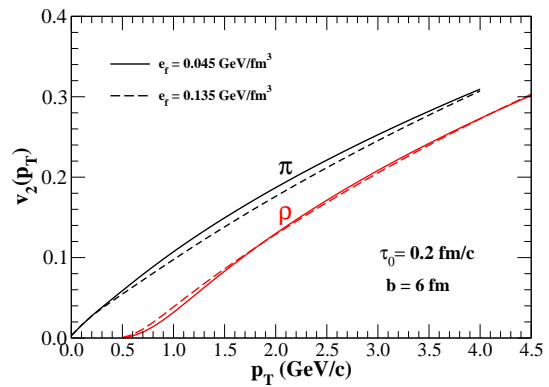,angle=-90,width=7.9cm}}
\caption{(Colour on-line) Effect of changing the freeze-out
energy density on the spectra (upper panel) and $v_2$ (lower panel)
for $\pi$ and $\rho$ mesons. The plot for $v_2$ for the $\rho$-mesons
is shifted by 0.5 GeV/$c$ along the x-axis.}
\label{fig3}
\end{figure}

The production of thermal photons is calculated by folding the history of
the evolution of the system with the rate for the production of photons
from the quark matter and the hadronic matter. We use the complete 
leading-order results for the production of photons from the QGP~\cite{guy}
 and the latest results for the radiation of photons from a hot hadronic 
gas~\cite{simon}. As mentioned earlier,  the equation of state 
(EOS Q~\cite{ksh}) incorporating a phase transition to quark gluon plasma 
at $T\approx$ 164 MeV, and  a resonance gas for the hadronic phase below 
the energy density of 0.45 GeV/fm$^{-3}$ is used to describe the evolution. 
The mixed phase is described using Maxwell's construction.  The freeze-out 
is assumed to take place at $e_f$ = 0.075 GeV/fm$^3$. The spectrum of 
hadrons is obtained using the Cooper-Fry formulation.

As a first step we show the  calculated spectra of primary pions and rho 
mesons along with the differential elliptic flow for a typical impact 
parameter of 6 fm for Au+Au collisions at RHIC energies. We see that if 
the initial time is large, the inverse slope of the pion spectrum 
(Fig.~\ref{fig1}; upper panel) is smaller due to a late start of flow 
(all the results are for zero rapidity). However the deviation in the spectrum 
up to about 1.5 GeV/$c$ is not substantial, even though the spectrum at 
$p_T \approx 4$ GeV/$c$ changes by about a factor of three. It is very 
interesting to note that the differential elliptic flow is almost 
independent of the initial time across the range of $p_T$ values 
considered. Similar results are obtained for the primary rho mesons, 
even though the change in the spectra is more noticeable as the rho 
meson (see Fig.~\ref{fig2}) being heavier is more strongly affected 
by the flow. We have verified that similar results are obtained for 
all other particles, included in the (nearly complete) list of hadrons 
in our calculations. The similarity of spectra for the primaries 
ensures that the spectra of particles after the resonance decays are 
accounted for, will remain similar.

One may argue that the freeze-out at a given (rather low) energy density 
represents an over-simplification of the decoupling of the hadrons from 
the interacting medium. Thus one could consider  hadrons decoupling at a 
sufficiently large temperature or energy density from the hydrodynamic 
evolution and populating a hadronic cascade from which they undergo a 
freeze-out depending on their interaction cross-sections~\cite{bass1,edward}. 
While a full calculation along these lines would be welcome, we can get a 
feeling of the outcome by estimating the effect of the freeze-out density 
on the spectra and the $v_2$ of the hadrons using hydrodynamics.

In Fig.~\ref{fig3}, we give our results for the $p_T$ distributions and 
differential elliptic flow, $v_2(p_T)$, for pions and $\rho$-mesons for 
energy-densities of 0.045 GeV/fm$^3$ and 0.135 GeV/fm$^3$, which 
correspond to freeze-out temperatures of about 100 and 140 MeV 
respectively. We note that the $p_T$ distribution continues to 
evolve as the system cools from 140 MeV to 100 MeV, due to the 
radial flow. However, the elliptic flow is seen to have essentially 
acquired its final value when the temperature is still large. Thus 
we feel that if these hadrons are fed into a hadronic cascade the 
results for the $v_2$ will remain unaltered.  

We conclude thus,  that it would not be easy to determine $\tau_0$ on 
the basis of either spectra or $v_2$ for hadrons. The calculated 
differences in the spectra at very large values of $p_T$ may
not be quite useful, as the hydrodynamic description itself may 
not hold at $p_T \gg$ 2 GeV/$c$.

\begin{figure}[ht]
\centerline{\epsfig{file=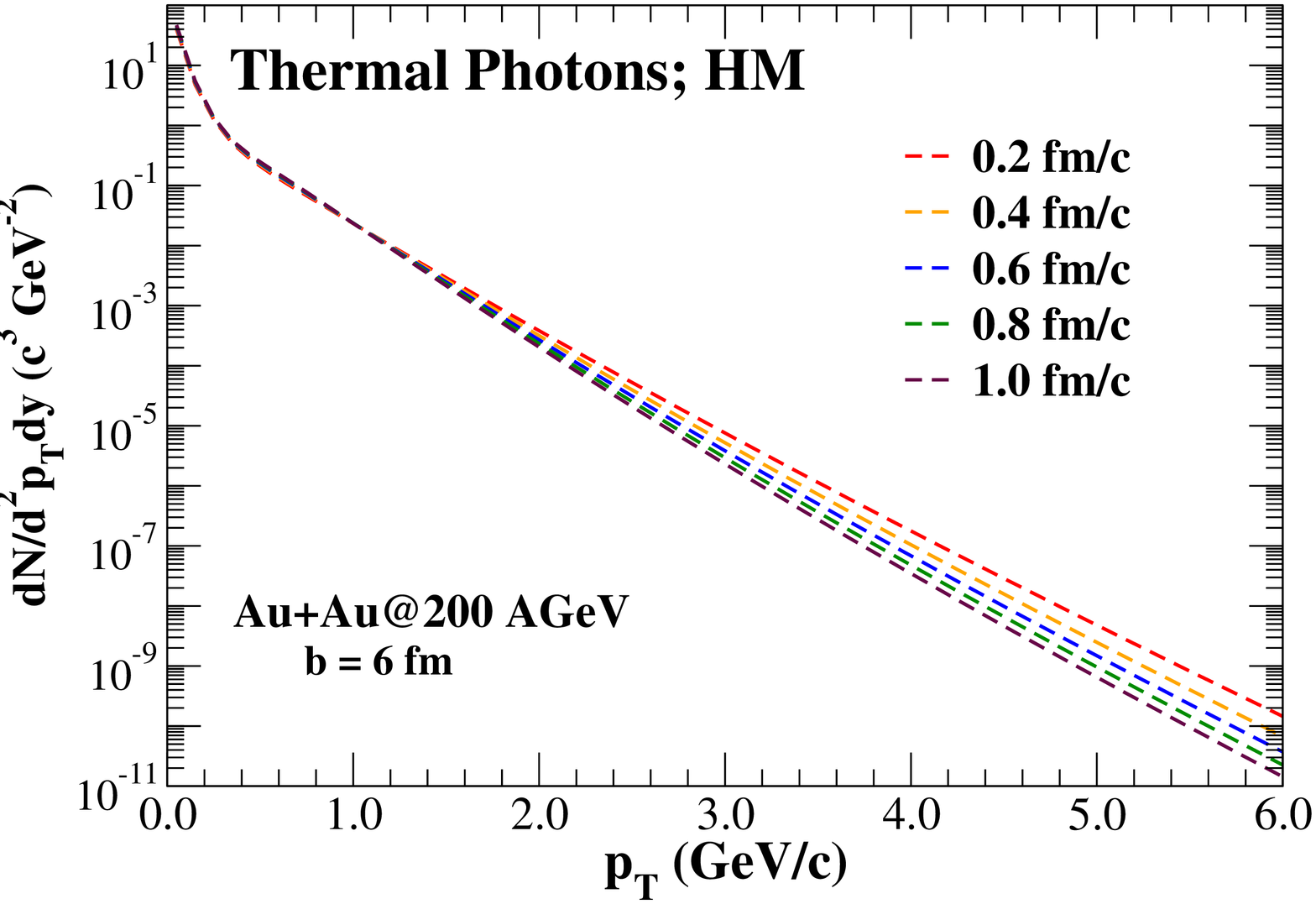,angle=-90,width=7.9cm}}
\centerline{\epsfig{file=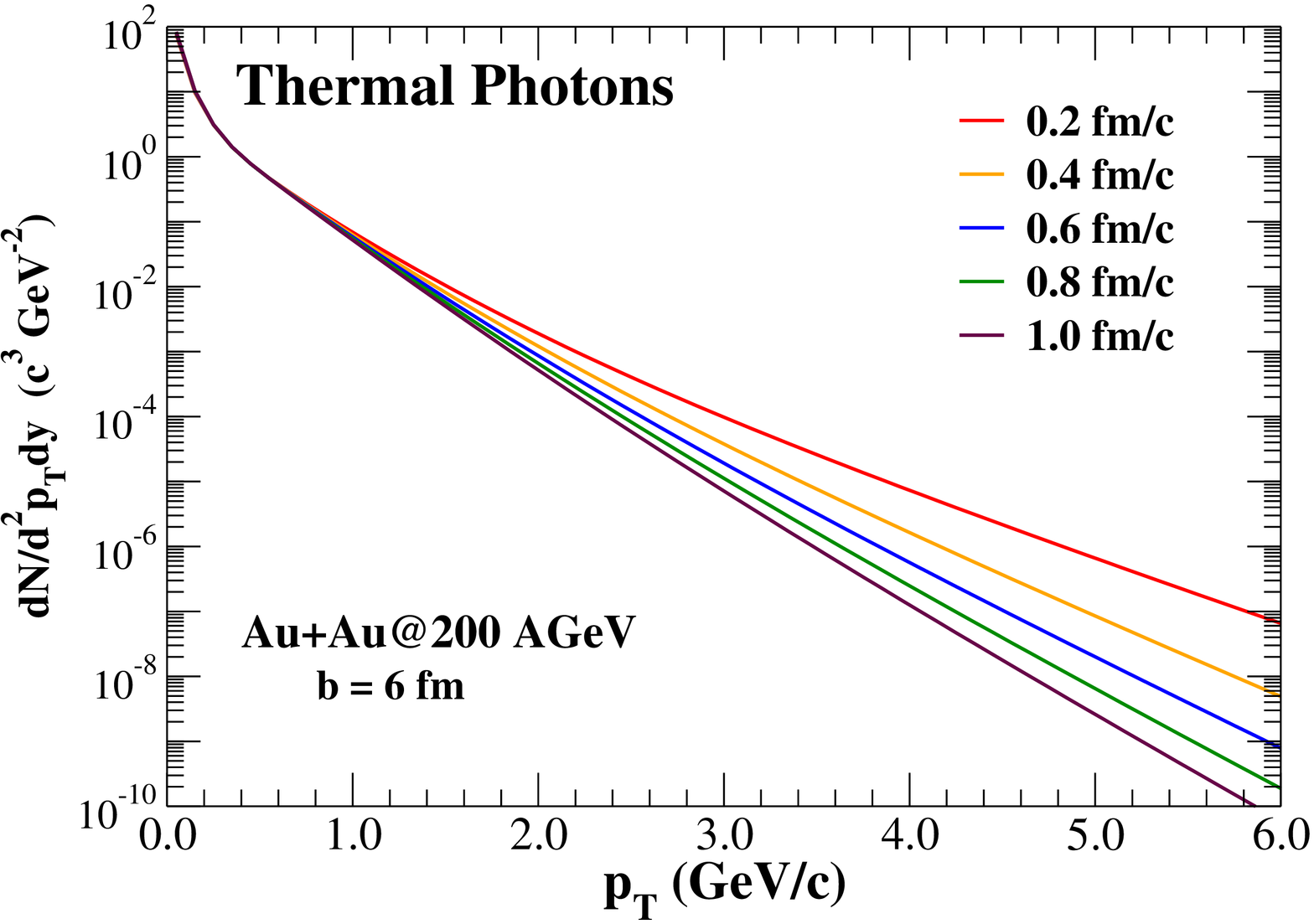,angle=-90,width=7.9cm}}
\caption{(Colour on-line) Spectrum of photons radiated from hadronic 
matter alone (upper panel) and the sum (lower panel) for different 
initial times.}
\label{fig4}
\end{figure} 

\begin{figure}[ht]
\centerline{\epsfig{file=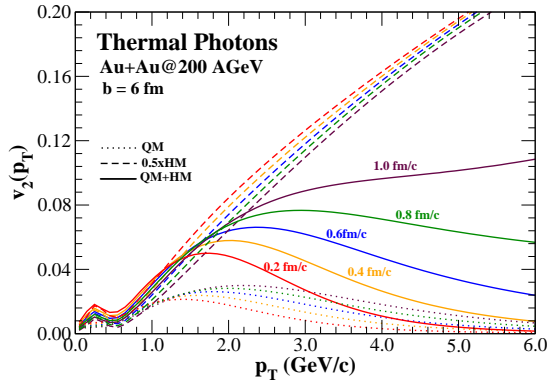,angle=-90,width=7.9cm}}
\caption{(Colour on-line) The differential elliptic flow of thermal 
photons (solid curves) for different initial time $\tau_0$. The results 
for emissions from the hadronic matter (long-dashed) and quark matter 
(short dashed) are also shown.}
\label{fig5}
\end{figure} 

\begin{figure}[ht]
\centerline{\epsfig{file=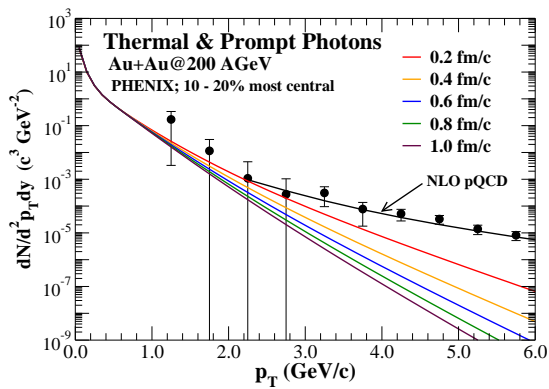,angle=-90,width=7.9cm}}
\caption{(Colour on-line) Single photons for 10--20\% most central 
collisions for Au+Au system at 200 GeV/A. NLO pQCD results for prompt 
photons (see text) and thermal photons for different formation times 
are also shown.}
\label{fig6}
\end{figure} 

Next we calculate the thermal photon production for the same system 
(Fig.~\ref{fig4}) using $\tau_0$ varying from 0.2 fm/$c$ to 1 fm/$c$. 
It is seen  that the contributions from the hadronic matter for different 
starting times  are only marginally different at lower $p_T$, though at 
6 GeV/$c$ it increases by an order of magnitude, as $\tau_0$ drops from 
1.0 fm/$c$ to 0.2 fm/$c$, due to  an increased flow for lower initial 
times. However, the smaller initial times  also admit much larger initial 
temperatures and the radiation of high $p_T$ photons from the quark matter 
increases substantially. Thus, while the total production of thermal 
photons for $p_T <$ 1 GeV/$c$ is not strongly affected, at $p_T $ = 6 
GeV/$c$, it increases by more than four orders of magnitude. This is 
known for quite some time in the literature.
 
The differential elliptical flow for thermal photons shows a much more
complex and rich structure (Fig.~\ref{fig5}). The $v_2$ for photons 
from the quark matter is much smaller than that for photons from the
hadronic matter~\cite{v2_phot}. This, happens as the anisotropies in 
momenta of the particles producing photons take some time to develop 
and the photons from the quark matter measure these anisotropies for 
early times. Note also the peak around 0.5 GeV/$c$ due to the dominating 
role of the reaction $\pi \rho \rightarrow \pi \gamma$ \cite{v2_phot} 
beyond about 0.5 GeV/$c$. We also see an interesting trend; the $v_2$ 
for thermal photons from the quark matter increases with the increase 
in the initial time $\tau_0$, while those for the photons from the 
hadronic matter  decrease with increase in the initial time $\tau_0$. 
This, we feel, happens due to a competition between larger initial 
(spatial) anisotropies and temperatures for smaller $\tau_0$ and a 
start of cooling and equalization of pressure gradients at an earlier 
time.

The results for the final $v_2$ for thermal photons (solid curves, 
Fig.~\ref{fig5}) reveal a large sensitivity to the initial time 
$\tau_0$, for $p_T$ greater than about 1.5 GeV/$c$. 
 This is brought about by  the increasing contribution of 
photons from quark-matter (with smaller $v_2$) as the time $\tau_0$ 
decreases. With increasing $\tau_0$ the relative contribution  of 
thermal photons from the hadronic matter increases. As these have 
larger $v_2$, the over-all $v_2$ becomes larger.

Before closing we show our results for single photon production and 
compare it with the measurements by the PHENIX experiment~\cite{phenix_phot} 
for the 10--20\% most central collisions for Au+Au system at 200 GeV/A 
(Fig.~\ref{fig6}). The prompt photon production has been calculated 
using NLO pQCD~\cite{pat} and choosing the factorization, renormalization, 
and fragmentation scales as equal to $p_T$/2. The effects of (impact 
parameter dependent) shadowing~\cite{eks98} and iso-spin is explicitly
accounted for. CTEQ5M structure functions are used. The thermal photon 
production for different initial times is also shown. We see that the 
NLO pQCD calculations nearly fully account for the production of photons 
at $p_T \geq$ 3 GeV/$c$.

We would like to add though that the fragmentation contribution to
the prompt photons (about 30\% for $p_T >$ 3 GeV) will be suppressed 
due to jet-quenching. Assuming a suppression by a factor of about 2 
(for quark jets which fragment into photons) this would mean a 
reduction by $\approx$ 15\% in the pQCD yield compared to what is 
shown here. However, this would be somewhat off-set by the production 
of photons due to passage of jets through the QGP~\cite{fms_phot}.
A more detailed and complete calculation will consider all these 
sources~\cite{v2_gale}. We postpone that to a future publication as 
the present emphasis is on the azimuthal anisotropy of thermal photons, 
which is rather negligible for other direct photons. Similarly,
one may consider the pre-equilibrium contribution as well
~\cite{bms_phot}, if $\tau_0$ is assumed to be large. These 
contributions will not show any $v_2$ as they are not subjected 
to collectivity.

If we are unable to separate these other sources of photons by, say,
either by tagging or by calculations, then the $v_2$ given here will 
have to be moderated as $(v_2^{\rm th}\times N_{\rm th}+v_2^{\rm non-th} 
\times N_{\rm non-th})/(N_{\rm th}+N_{\rm non-th})$, where `th' stands 
for `thermal'.  However as $v_2^{\rm non-th}$ is quite small, the 
difference in the resulting $v_2$ for single photons will still 
persist, being large for large $\tau_0$. In this connection, the 
success of the NLO pQCD in predicting the prompt contribution can 
be used with a great advantage as it can be subtracted from the 
direct photons to greatly increase the sensitivity of the remainder 
to $\tau_0$.

In order to get a more quantitative idea about these, we draw the 
attention of the readers to FIG.6 of Ref.~\cite{v2_gale}. First of 
all, as remarked earlier, the $Compton + annihilation$ part of the 
prompt photons does not contribute to azimuthal anisotropy. The 
prompt-fragmentation part, again as suggested earlier, will show 
a marginal positive $v_2$, due to the reaction plane dependence 
of energy loss suffered by the out-going quarks before fragmenting. 
This according to Ref.~\cite{v2_gale} is less than about $+1$\%. 
The jet conversion photons give rise to very small (less than $-1$\%) 
$v_2$. The resulting $v_2$ due to these two processes is essentially 
zero~\cite{v2_gale}. This, along with the possibility of an accurate 
determination of prompt photons using NLO pQCD as seen here holds out 
the promise of determining the sensitivity of elliptic flow of 
thermal photons to $\tau_0$.

An experimental verification of these predictions will give a very 
useful information about the formation time and also validate our ideas 
about the evolution of the hot and dense system in relativistic heavy 
ion collisions. A theoretical reanalysis of single photon production 
for Pb+Pb collisions at the CERN SPS has revealed similar sensitivity 
to the formation time~\cite{rupa}. 

In brief, we have calculated the thermal photon production for non-central 
collisions of gold nuclei at RHIC energies. The initial time is varied 
from 0.2 fm/$c$ to 1 fm/$c$, keeping the total entropy fixed and assuming 
formation of quark gluon plasma in thermal and chemical equilibrium, 
undergoing expansion and cooling. The $v_2(p_T)$ for hadrons is seen to 
be only marginally affected by the variations in the formation time. The 
spectra as well as the $v_2(p_T)$ of thermal photons are seen to be 
strongly affected by the formation time, as smaller initial times admit 
larger initial temperatures and larger contributions from the QGP phase,
when the momentum anisotropy is not fully developed. This could prove 
useful in determining the formation time $\tau_0$.

We thank Sangyong Jeon for his help in setting up the prompt photon 
calculation program of Patrick Aurenche's group.


\begin{thebibliography}{99}

\bibitem{jetq_theo}
X.~N.~Wang, Phys.\ Rev.\ C {\bf 63}, 054902 (2001);
M.~Gyulassy, I.~Vitev, X.~N.~Wang, Phys.\ Rev.\ Lett.\ {\bf 86}, 2537 (2001).

\bibitem{jetq_exp}
K.~Adcox {\it et al.} [PHENIX Collaboration], Phys.\ Rev.\ Lett. {\bf 88},
022301 (2002);
J.~Adams {\it et al.}  [STAR Collaboration],
Phys.\ Rev.\ Lett.\  {\bf 91}, 172302 (2003).

\bibitem{v2_theo}
P.~Huovinen, P.~F.~Kolb, U.~W.~Heinz, P.~V.~Ruuskanen and  S.~A.~Voloshin,
Phys.\ Lett.\ B {\bf 503}, 58 (2001).

\bibitem{v2_exp}
S.~S.~Adler {\it et al.} [PHENIX Collaboration],
Phys.\ Rev.\ Lett.\  {\bf 91}, 182301 (2003);
S.~Esumi (for the PHENIX collaboration),
Nucl.\ Phys.\  A{\bf 715}, 599 (2003);
C.~Adler {\it et al.} [STAR Collaboration], Phys.\ Rev.\ Lett.\ {\bf 90},
032301 (2003); {\it ibid.} {\bf 89} 132301 (2002); {\it ibid.}
{\bf 87} 182301 (2001).


\bibitem{phenix_phot} S.~S.~Adler {\it et al.} [PHENIX Collaboration], Phys. \
Rev. \ Lett. {\bf 94}, 232301 (2005).


\bibitem{kms} J. Kapusta, L. McLerran, and D. K. Srivastava,
\ Phys. \ Lett. \ B {\bf 283}, 145 (1992).

\bibitem{biro} T. S. Biro, E. van Doorn, B. M\"uller, M. H. Thoma,
and X. N. Wang, \ Phys. \ Rev. \ C {\bf 48}, 1275 (1993).

\bibitem{kkg} K. Geiger and B. M\"uller, \ Nucl. \ Phys.\ B {\bf 369}, 600 (1992).

\bibitem{bms} S. A. Bass, B. M\"uller, and D. K. Srivastava, 
\ Phys. \ Lett. \ B {\bf 551}, 277 (2003).

\bibitem{xu} Zhe Xu and C. Greiner, \ Phys. \ Rev. \ C {\bf 71}, 064901 (2005).

\bibitem{sspc} K. Eskola, B. M\"uller, and X.-N. Wang, \ Phys. \ Lett. \ B
{\bf 374}, 20 (1996).

\bibitem{eskola} K. J. Eskola, K. Kajantie, P. V. Ruuskanen, and K. Tuominen,
 \ Nucl. \ Phys. \ B {\bf 570}, 379 (2000).

\bibitem{son} R. Baier, A. H. Mueller, D. Schiff, and D. T. Son,
\ Phys. \ Lett. \ B {\bf 502}, 51 (2001).

\bibitem{munshi} 
 D. K. Srivastava, M. G. Mustafa, and B. M\"uller,
\ Phys. \ Rev. \ C {\bf 56}, 1064 (1997);
D. M. Elliot and D. H. Rischke, \ Nucl. \ Phys. \ A {\bf 671},
583 (2000).

\bibitem{stan} S. Mrowczynski, \ Phys. \ Lett. \ B {\bf 214}, 587 (1988).

\bibitem{guy_inst} P. Arnold, J. Lanaghan, and G. D. Moore, JHEP {\bf 0308}, 002 
(2003).

\bibitem{v2_phot} R. Chatterjee, E. S. Frodermann, U. W. Heinz,
and D. K. Srivastava, \ Phys. \ Rev. \ Lett. {\bf 96}, 202302 (2006).

\bibitem{v2_dil} R. Chatterjee, D. K. Srivastava, U. W. Heinz,
and C. Gale, \ Phys. \ Rev. \ C {\bf 75}, 054909 (2007); U. W. Heinz, 
R. Chatterjee, E. Frodermann, C. Gale, and D. K. Srivastava,
\ Nucl. \ Phys. \ A {\bf 783}, 379 (2007).

\bibitem{bass} R. J. Fries, B. M\"uller, C. Nonaka,  and S. A. Bass, \ Phys. \ Rev.
\ Lett. {\bf 90}, 202303 (2003);
R. C. Hwa and C. B. Yang, \ Phys. \ Rev. \ C {\bf 67}, 064902 (2003).

\bibitem{uk} U. Heinz and A. Kuhlman, \ Phys. \ Rev. \ Lett. {\bf 94}, 132301 (2003).

\bibitem{bj} J. D. Bjorken, \ Phys. \ Rev. \ D {\bf 27}, 140 (1983).

\bibitem{ksh} P.~F.~Kolb, J.~Sollfrank, and U.~Heinz, \ Phys. \  Rev. \ C
 {\bf 62}, 054909 (2000).

\bibitem{guy} P.~Arnold, G.~D.~Moore, and L.~G.~Yaffe, JHEP {\bf 0112}, 009 
(2001).

\bibitem{simon} S.~Turbide, R.~Rapp, and C.~Gale, \ Phys. \ Rev. \ C {\bf 69},
 014903 (2004).

\bibitem{bass1} S. A. Bass and A. Dumitru, \ Phys. \ Rev. \ C {\bf 61}, 064909 
(2000).

\bibitem{edward} D. Teaney, J. Lauret, and E. V. Shuryak, 
arXiv:nucl-th/0110037.

\bibitem{pat} P. Aurenche, M. Fontannaz, J.-P. Guillet, B. A. Knielhl,
E. Pilon, and M. Werlen, \ Eur. \ Phys. \ Jour. \ C {\bf 9}, 107 (1999).

\bibitem{eks98} K. J. Eskola, V. J. Kolhinen, and C. A. Salgado,
 \ Eur. \ Phys. \ Jour. \ C {\bf 9}, 61 (1991).


\bibitem{fms_phot} R.~J.~Fries, B.~M\"uller, and D.~K.~Srivastava,
Phys. \ Rev. \ Lett. {\bf 90}, 132301 (2003); R.~J.~Fries, B.~M\"uller,
and D.~K.~Srivastava, Phys. \ Rev. \ C {\bf 72} 041902 (R) (2005).

\bibitem{v2_gale} S. Turbide, C. Gale, E. Frodermann, and U. Heinz,
\ Phys. \ Rev. \ C {\bf 77}, 024909 (2008).

\bibitem{bms_phot} S.~A.~Bass, B.~M\"uller, and D.~K.~Srivastava,
Phys. \ Rev. \ Lett. {\bf 90}, 082301 (2003);
 T.~ Renk, S.~A.~Bass, and D.~K.~
Srivastava, Phys. \ Lett. B {\bf 632}, 632 (2006).

\bibitem{rupa} R. Chatterjee, D. K. Srivastava, S. Jeon, and C. Gale, to 
be published.

\end{thebibliography}
\end{document}